\documentclass[aps,prc,twocolumn,groupedaddress]{revtex4-1}
\usepackage{graphics}
\usepackage{epsfig}
\usepackage{graphicx}
\usepackage{longtable}
\usepackage{amssymb}
\begin{document}

% Use the \preprint command to place your local institutional report
% number in the upper righthand corner of the title page in preprint mode.
% Multiple \preprint commands are allowed.
% Use the 'preprintnumbers' class option to override journal defaults
% to display numbers if necessary
%\preprint{}

%Title of paper
\title{Penning-Trap Mass Measurements of Neutron-Rich K and Ca Isotopes:\\ Resurgence of the $N=28$ Shell Strength}

% repeat the \author .. \affiliation  etc. as needed
% \email, \thanks, \homepage, \altaffiliation all apply to the current
% author. Explanatory text should go in the []'s, actual e-mail
% address or url should go in the {}'s for \email and \homepage.
% Please use the appropriate macro foreach each type of information

% \affiliation command applies to all authors since the last
% \affiliation command. The \affiliation command should follow the
% other information
% \affiliation can be followed by \email, \homepage, \thanks as well.
\author{A. Lapierre$^{1, 2}$, M. Brodeur$^{1, 3}$, T. Brunner$^{1,4}$, S. Ettenauer$^{1, 3}$, P. Finlay$^5$, \\ A. T. Gallant$^{1, 3}$, V. V. Simon$^{1, 7, 8}$, P. Delheij$^1$, D. Lunney$^9$, R. Ringle$^2$, H. Savajols$^6$, and J. Dilling$^1$}
%\email[]{Your e-mail address}
%\homepage[]{Your web page}
%\thanks{}
%\altaffiliation{}
\affiliation{$^1$ TRIUMF, 4004 Wesbrook Mall, Vancouver, BC, V6T 2A3, Canada}
\affiliation{$^2$ NSCL, Michigan State University, East Lansing, MI, 48824, USA}
\affiliation{$^3$ Department of Physics and Astronomy, University of British Columbia, Vancouver, BC, V6T 1Z1, Canada}
\affiliation{$^4$ Physik Department E12, Technische Universit\"{a}t M\"{u}nchen, James Franck Str., Garching, Germany}
\affiliation{$^5$ Department of Physics, University of Guelph, Guelph, ON, N1G 2W1, Canada}
\affiliation{$^6$ GANIL, Blvd Henri Becquerel, BP 55027-14076, CAEN, Cedex 05, France}
\affiliation{$^7$ Max-Planck-Institut f\"{u}r Kernphysik, Saupfercheckweg 1, 69117, Heidelberg, Germany}
\affiliation{$^8$ University of Heidelberg, 69117, Heidelberg, Germany}
\affiliation{$^9$ CSNSM-IN2P3-CNRS, Universit\'{e} de Paris Sud, 91405 Orsay, France}

%Collaboration name if desired (requires use of superscriptaddress
%option in \documentclass). \noaffiliation is required (may also be
%used with the \author command).
%\collaboration can be followed by \email, \homepage, \thanks as well.
%\collaboration{}
%\noaffiliation

\date{\today}

\begin{abstract}
We present Penning-trap mass measurements of neutron-rich $^{44, 47-50}$K and $^{49, 50}$Ca isotopes carried out at the TITAN facility at TRIUMF-ISAC. The $^{44}$K mass measurement was performed with a charge-bred 4+ ion utilizing the TITAN EBIT, and agrees with the literature. The mass excesses obtained for $^{47}$K and $^{49, 50}$Ca are more precise and agree with the values published in the 2003 Atomic Mass Evaluation (AME'03). The $^{48, 49, 50}$K mass excesses are more precise than the AME'03 values by more than one order of magnitude. For $^{48, 49}$K, we find deviations by 7$\sigma$ and 10$\sigma$, respectively. The new $^{49}$K mass excess lowers significantly the two-neutron separation energy at the neutron number $N=30$ compared with the separation energy calculated from the AME'03 mass-excess values, and thus, increases the $N=28$ neutron-shell gap energy at $Z=19$ by approximately 1 MeV.
\end{abstract}

% insert suggested PACS numbers in braces on next line
% insert suggested keywords - APS authors don't need to do this
%\keywords{}

%\maketitle must follow title, authors, abstract, \pacs, and \keywords
\maketitle

% body of paper here - Use proper section commands
% References should be done using the \cite, \ref, and \label commands
% Put \label in argument of \section for cross-referencing
%\section{\label{}}
\section{Introduction}
The magic numbers associated with the closure of nuclear shells are amongst the most fundamental quantities governing the nuclear structure \cite{otsu01, krue10}. Their disappearance away from the valley of stability holds great intrigue for theories that seek to correctly describe the nuclear interaction \cite{sorl08}. The nuclear binding energy is sensitive to the existence of shell structures, and hence, precision mass measurements can provide signatures of shell structure modifications as well as new magic numbers \cite{mittig97, lunn03}. The original case study for the disappearance of a magic number was the study of $N=20$ and the `Island of Inversion' in neutron-rich nuclei of atomic numbers $Z\sim10-12$ \cite{war90, camp75}, which were discovered from pioneering online mass spectrometry of Na isotopes \cite{thib75}. Nuclear spectroscopy \cite{detraz79, detraz83} and subsequent work have revealed that the extra binding energy giving unexpected stability to these nuclei, is the result of a deformation caused by the inversion of so-called `intruding' $pf$ orbitals, as originally hypothesized in Ref. \cite{war90}. There are still presently intensive efforts underway to delineate the island's shore and understand the role of the interplay between spherical and deformed configurations on the nuclear stability in this mass region. An excellent example of such work is the recent discovery of a coexisting 0+ state in $^{32}$Mg reported by Wimmer \textit{et al.} \cite{wimm10}. 

As a natural extension, attention has turned to the spin-orbit closed-shell $N=28$ and its own `Island of Inversion'. The erosion of the $N=28$ shell was observed from the determination of the lifetime and deformation of $^{44}$S \cite{sorlin93, glas97}. The weakening of the $N=28$ neutron-shell gap was observed in mass measurements of neutron-rich Si, P, S, Cl isotopes \cite{sava05, sara00, jura07, ring09}, where an isomer in $^{43}$S has shed light on shape coexistence in this mass region \cite{sara00}. Deformation develops gradually from the spherical $^{48}$Ca to the deformed $^{42}$Si \cite{bast07}. Between these two extremes, the spherical and deformed shapes compete, as shown recently in $^{44}$S \cite{force10}. Further experimental data in this $Z\sim14-17$ region are needed to understand the shell-breaking mechanism of the $N=28$ spin-orbit shell closure, which is also of importance to understand the evolution of other shell gaps having the same origin.

In this article, we report Penning-trap mass measurements at TITAN of neutron-rich $^{44, 47-50}$K isotopes from which we obtain a $N=28$ neutron-shell gap energy dramatically stronger than the previously determined value deduced from the AME'03 mass excesses. We also report mass measurements for the isotopes $^{49, 50}$Ca. Uncertainties have been reduced between one and two orders of magnitude and reveal major deviations from previous values, determined from $\beta$-decay and reaction studies.

\begin{table*}[t]
\caption{Frequency ratios (relative to stable $^{39}$K) and atomic mass-excess (ME) values of the investigated K and Ca isotopes. The first displayed uncertainty is the statistical error while the second one is the systematic error (see text), the third error in square brackets is the total uncertainty which is the result of the statistical and systematic errors added in quadrature. $\delta_{ME}$ is the deviation of the AME'03 mass-excess values \cite{AME03_1, AME03_2} with respect to our results ($ME_{expt}-ME_{AME'03}$). \label{tab:CaKresults}}
\begin{ruledtabular}
\begin{tabular}{c c c c c c}
Isotopes              & Half-life & $\nu^{meas}_{c}/\nu^{ref}_c$ & $ME_{expt}$ (keV)    & $ME_{AME'03}$ (keV) & $\delta_{ME}$ \\
\hline
$^{44}$K$^{4+}_{25} $ & 22.13(19) m         &  0.886306820(35)(0.5)[35]     &  -35778.7(1.6)(0.02)[1.6]    & -35810(36)         &  31(36)                         \\

$^{47}$K$^{+}_{28}$ & 17.50(24) s        &  0.829689831(27)(3)[27]       &  -35711.8(1.4)(0.2)[1.4]     & -35696(8)          &  -15(8)                         \\
    
$^{48}$K$^{+}_{29}$ & 6.8(2) s      &   0.812328289(14)(4)[14]   &  -32284.2(0.8)(0.2)[0.8]    & -32124(24)         &   -160(24)                  \\

$^{49}$K$^{+}_{30}$ & 1.26(5) s     &   0.795691609(13)(4)[14]     &  -29611.3(0.8)(0.2)[0.8]   & -30320(70)         &   708(70)                      \\

$^{50}$K$^{+}_{31}$ & 472(4) ms   &   0.779702420(130)(4)[130]     &  -25727.6(7.7)(0.3)[7.7]         & -25352(278)        &   -376(278)                      \\ 

$^{49}$Ca$^{+}_{29}$ & 8.718(5) m  &   0.795895563(20)(4)[20]    &  -41300.0(1.2)(0.2)[1.2]        & -41289(4)          &   -11(4)                       \\ 

$^{50}$Ca$^{+}_{30}$ & 13.9(6) s         &   0.779934672(26)(4)[26]     &  -39589.0(1.6)(0.3)[1.6]      & -39571(9)          &   -18(9)                        \\
\end{tabular}
\end{ruledtabular}
\end{table*}

\section{Experimental set-up and method}
The measurements were conducted at the TITAN (TRIUMF's Ion Traps for Atomic and Nuclear science) facility \cite{dill03, dehl06} located at the TRIUMF-ISAC exotic nuclide facility. Over the last few years, TITAN has been successfully used to measure masses of neutron-rich isotopes with short half-lives, such as $^{6,8}$He \cite{ryjk08, max12}, $^{11}$Li \cite{smit08}, and $^{11, 12}$Be \cite{ring08, ette10}. Recently, TITAN also demonstated the first use of high charge states for mass measurements of short-lived nuclides in a Penning trap \cite{ette11}. TITAN is presently composed of three ion traps: a linear radiofrequency quadrupole (RFCT) cooler and buncher \cite{RFQCT}, an electron beam ion trap (EBIT) charge breeder \cite{froe06, lapi10}, and a 3.7-T Penning-trap mass spectrometer \cite{max09, max10, max11}. The set-up is also equipped with a surface ion source located under the TITAN RFCT to supply ions of stable alkali isotopes such as $^{39}$K for systematic tests, optimization, and for mass calibration.

The short-lived, neutron-rich K and Ca isotope beams were produced by TRIUMF's ISAC (Isotope Separator and ACcelerator) radioactive beam facility \cite{Domb00} with a surface ion source using a Ta target bombarded by a $\sim$75-$\mu$A (and later reduced to $\sim$40 $\mu$A) 500-MeV proton beam. The ISAC beams were mass separated with a dipole magnet with a mass resolving power of $\sim$3000 and delivered to TITAN with a kinetic energy of 15 keV. The singly charged ion beams were injected into the RFCT where they were decelerated electrostatically and their transverse and longitudinal emittances were reduced with He buffer gas. The ions were subsequently extracted as bunches with an energy of approximately 2 keV. The bunches were then either sent directly to the Penning trap, or to the EBIT for charge breeding and subsequently to the Penning trap. During this experiment, the EBIT was utilized only for $^{44}$K (T$_{1/2}$=22.13 m) to achieve a charge state of $4+$ and proof the principle of using such a breeder for mass measurements of radioactive isotopes. $^{44}$K$^{4+}$ was charge bred with an electron-beam energy of 3.95 keV, and a weak electron-beam current of less than 1 mA produced by only warming up the Pierce-type electron-gun cathode. The EBIT magnetic-field strength was 4 T, the trapping potential was set to 100 V, and the charge breeding time was 200 ms.

\begin{figure}[t]
  \centering
  \includegraphics*[width=7.5cm]{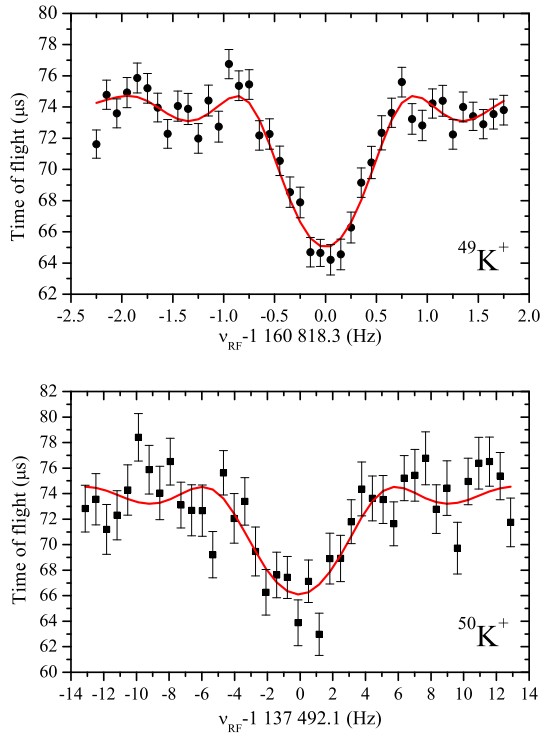}
  \caption{(Color online) $^{49}$K$^{+}$ and $^{50}$K$^{+}$ TOF-ICR resonance curves. The RF quadrupolar excitation times were 997 and 147 ms, respectively. The red solid curve is a fit of the theoretically expected line shape \cite{blaum06} to the data points.}
  \label{fig:K4950rescurv}
\end{figure}

The mass measurements were performed using the well-established time-of-flight ion cyclotron resonance (TOF-ICR) technique \cite{blaum06}. For details of the TITAN measurement procedure see Refs. \cite{max09, max10, max11}. Quadrupole RF excitation times ranged overall from 8 to 997 ms, but excitation times of 147 and 997 ms were normally used. Short excitation times were used for setting up the trap parameters such as the excitation frequency range, while overall a 997 ms excitation time was chosen for continuous high-precision data taking. Prior to quadrupole RF excitation, a dipole RF field was applied to expel out of the trap isobaric contaminating ions such as Cr$^{+}$ an Ca$^{+}$. Dipole excitation ranged from a few to tens of ms. After quadrupole excitation of the studied trapped K and Ca ions, they were ejected from the trap and the energy gained during the excitation was adiabatically converted in the decreasing magnetic field gradient into the axial energy, which resulted in shorter time of flights to a microchannel plate (MCP) detector. Full conversion of magnetron motion into cyclotron motion only occured when the RF frequency was equal to the ions' cyclotron frequency, $\nu_c$, which was thus determined by scanning the frequency of the RF field. In addition, throughout the experiment, the number of trapped ions was limited to $1-2$ per frequency measurement step to avoid frequency shifts due to ion-ion interaction \cite{max09, max10, max11}. Typical TOF-ICR curves obtained with $^{49}$K$^{+}$ and $^{50}$K$^{+}$ are shown in Fig. \ref{fig:K4950rescurv}. The ions' mass was determined from the relation $\nu_c=qB/(2\pi m)$, where $q$ is the charge of the trapped ions, $m$ their mass, and $B$, the trap's magnetic field strength. $^{39}$K from the TITAN ion source was used as a mass reference to calibrate the trap's magnetic field strength. TOF-ICR measurements of $^{39}$K were taken before and after the resonance frequency measurements of the neutron-rich K and Ca isotopes.

\begin{figure}[t]
  \centering
\includegraphics*[width=8.9cm]{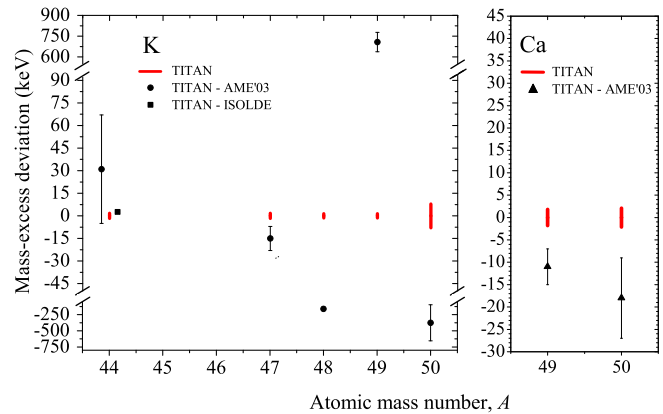}
\caption{(Color online) Deviation of the measured mass-excesses with respect to those published in AME'03. Our $^{44}$K mass excess of -35778.7(1.6) keV measured with multiply charged K$^{4+}$ agrees with the more precise ISOLTRAP value of -35781.29(0.47) keV \cite{yazi07}. The black error bars represent the uncertainties of the AME'03 values. The red thick elongated lines lying on the zero-deviation-energy line represent the uncertainties of our measurements.}
\label{fig:CaKresults}
\end{figure}

\section{Results}
The measured frequency ratios of the studied K and Ca isotopes with respect to stable $^{39}$K are listed in Table \ref{tab:CaKresults} along with their respective deduced mass-excess values. The frequency ratios are obtained from the weighted average of several frequency measurements, each of them resulting from a series of 50 to 200 frequency scans conducted with 41 frequency steps. Table \ref{tab:CaKresults} also presents the difference between the TITAN's mass-excess values and those in the AME'03 \cite{AME03_1, AME03_2}, $\delta_{ME}$. Fig. \ref{fig:CaKresults} shows the $\delta_{ME}$ deviation for the measured K and Ca isotopes.

The most significant sources of uncertainty are trap imperfections \cite{bro86}, such as misalignments and harmonic distortions as well as higher-order electric field multipoles. Ion-ion interactions, relativisitic effects, and magnetic-field decay can also cause frequency shifts. Systematic studies using stable species were performed for the TITAN system in order to evaluate such sources of systematic errors and were discussed in details in Refs. \cite{max09, max10, max11}. For a well-tuned trap, at the level of statistical precision of the present measurements, the principal systematic error affecting our measurements is due to magnetic field inhomogeneities, the misalignment of the trap electrodes with the magnetic field, the harmonic distortion and the non-harmonic terms in the trapping potential. In Ref. \cite{max11}, it was estimated that the total systematic error is $\pm$0.2(2) ppb per difference of atomic mass number over charge ($A/q$) between the investigated and reference ions. In Table \ref{tab:CaKresults}, a conservative error of 0.4 ppb is added in quadrature to the statistical error to obtain the total uncertainty, which is exclusively dominated by statistics.

The $^{44}$K mass excess was determined using $^{44}$K$^{4+}$. A preliminary result was recently presented in Ref. \cite{lapi10} along with a typical $^{44}$K$^{4+}$ resonance curve and further information on the measurement. Our final result of -35778.7(1.6) keV is in accordance with the AME'03 value, but approximately one order of magnitude more precise. It agrees well with the recent ISOLTRAP value of -35781.29(0.47) keV, which is a factor of 4 more accurate than ours \cite{yazi07}. The sum of the ionization potentials from neutral K to K$^{3+}$ is 143 eV \cite{nist} and was included in the calculations of the $^{44}$K mass excess. Note that this correction, however, is one order of magnitude smaller than the statistical error of 1.6 keV. The uncertainty of the $^{44}$K$^{4+}$ mass-excess measurement is comparable to the uncertainties of the measurements performed with singly charged ions because for this particular measurement, the quadrupole RF excitation time was kept short at 147 ms to avoid significant ion losses caused by charge-exchange recombination in the Penning trap. The uncertainty of a mass measurement is inversely proportional to the ion charge, excitation time, and square root of the detected number of ions. Hence, the increase in precision obtained by using an ion charge of ${4+}$ and a higher number of detected ions per second ($\sim \sqrt{5}=2.2$) was reduced by a factor of approximately 7 due to the use of a shorter RF excitation time. Our mass-excess value for $^{47}$K is in agreement, within 2$\sigma$, with the AME'03 value, which is based on three transfer-reaction measurements: $^{48}$Ca(d,$^3$He)$^{47}$K \cite{newm66} and $^{48}$Ca(t,$\alpha$)$^{47}$K \cite{will66, stoc68}. However, those obtained for $^{48, 49}$K show large deviations from the evaluated values by $\delta_{ME}=$-160(24) keV and $\delta_{ME}=$708(70) keV, respectively. The $^{48}$K mass-excess value in AME'03 is mainly infered from two transfer-reaction measurements: $^{48}$Ca($^7$Li,$^7$Be)$^{48}$K \cite{weis78} and $^{48}$Ca($^{14}$C,$^{14}$N)$^{48}$K \cite{maye80}. The $^{49}$K mass-excess value in AME'03 was mainly determined from a single measurement of the $\beta$-decay end-point energy \cite{mieh86}. The deviation of the AME'03 $^{50}$K mass excess from our measurement is within 2$\sigma$. The $^{50}$K mass excess in AME'03 is based on a direct measurement (Time-Of-Flight Isochronous (TOFI) spectrometer) \cite{tu90} as well as a determination of the $\beta$-decay end-point energy \cite{mieh86}. Our mass excesses for $^{49,50}$Ca are within 2$\sigma$ of the evaluated values. The $^{49}$Ca mass excess in AME'03 is deduced from three $^{48}$Ca(n,$\gamma$)$^{49}$Ca radiative neutron capture measurements \cite{arne69,cran70,AME03_1, AME03_2}, while the $^{50}$Ca excess value is based on two $^{48}$Ca(t,p)$^{50}$Ca measurements \cite{hind66, will66}.

\section{Discussion of the results}
\subsection{Disagreement of the $^{48}$K mass excess} 
The $^{48}$K mass excess in AME'03, which is obtained from transfer-reaction studies \cite{weis78,maye80}, is larger than our result by 160(24) keV. One possible explanation for this disagreement is the transfer-reaction studies might have measured the mass excess of an excited state and not the ground state. Based on two series of $\gamma$-spectroscopy coincidence measurements, Kr\'{o}las \textit{et al.} \cite{krol07} have recently proposed a new excited-state structure for $^{48}$K, whose first excited levels are at 143 keV (J$^{\pi}=$2-), 279 keV (J$^{\pi}=$2-), 728 keV (J$^{\pi}=$3-), and 2.117 MeV (J$^{\pi}=$5+). No peak at 143 keV was discerned in the transfer-reaction studies. Surprisingly, the energy of the 143-keV level is within the uncertainty of the 160(24)-keV deviation of our $^{48}$K mass-excess result. Note that Kr\'{o}las \textit{et al.} also proposed a new J$^{\pi}$=1- spin-parity assignment for the $^{48}$K ground state, which is consistent with the spin-parity evolution of odd K isotopes near $N=28$ \cite{krol07} and a recent spin-parity re-assignement of the $^{50}$K ground state as J$^{\pi}$=1- \cite{craw09}.

\begin{figure}[t]
  \centering
\includegraphics*[width=8.8cm]{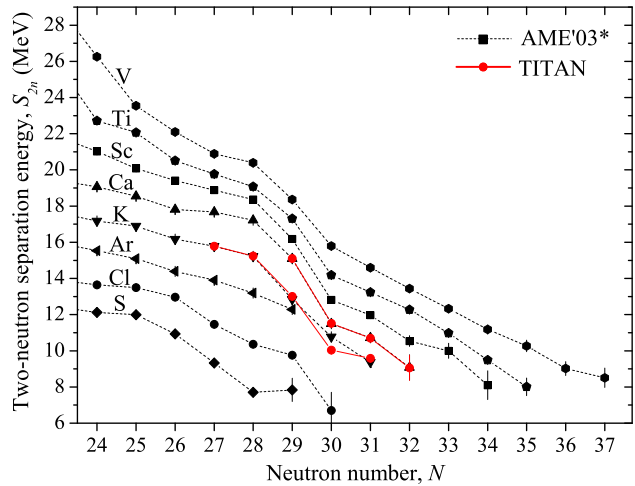}
\caption{(Color online) Two-neutron separation energy curves of neutron-rich S, Cl, Ar, K, Ca, Sc, Ti, and V isotopes. The red solid line shows the $S_{2n}$ values calculated with the results of our K and Ca mass measurements. The black dotted line, named AME'03*, represents those calculated from the AME'03 mass-excess values for which measured mass excesses of K \cite{yazi07}, S \cite{ring09, jura07}, and Cl \cite{jura07} isotopes were included with a weighted average.}
\label{fig:2nsep}
\end{figure}

\begin{figure}[t]
  \centering
\includegraphics*[width=8.5cm]{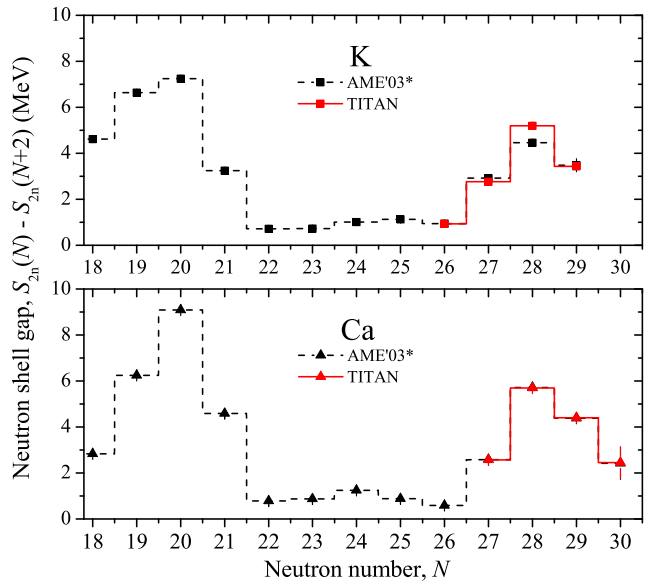}
\caption{(Color online) K and Ca neutron-shell gap energies as function of the neutron number $N$. Our K measurements increase significantly the value of the neutron-shell gap energies at $N=28$, with respect to the shell gap energy calculated from the AME'03 mass-excess values. \label{fig:KCashellgap}}
\end{figure}

\begin{figure}[t]
  \centering
\includegraphics*[width=8.8cm]{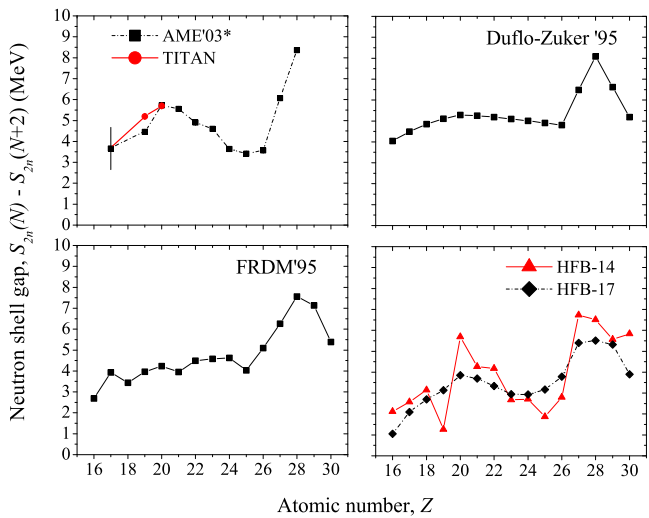}
\caption{(Color online) Experimental neutron-shell gap energy as a function of the atomic number $Z$ at $N=28$ compared with the shell gap energy obtained from various mass models (see text). \label{fig:K_SGvsZ}}
\end{figure}

\subsection{Separation and shell gap energies}
Structures in the nuclear shells can manifest themselves through the two-neutron separation energy, $S_{2n}(Z,N) = M(Z,N-2) + 2\cdot M_n - M(Z,N)$ as well as the neutron shell gap, $\Delta_{n}(Z,N)=S_{2n}(N)-S_{2n}(N+2)$, where $M(Z,N)$ is the atomic mass of a nucleus of atomic number $Z$ and $M_n$ is the neutron mass. Fig. \ref{fig:2nsep} presents the two-neutron separation energy curves of S to V isotopes, and compares the K and Ca curves obtained from our results with those calculated from the AME'03 mass-excess values to which we added the results of recent measurements (see caption) \cite{yazi07, ring09, jura07}. In Fig. \ref{fig:KCashellgap}, the K and Ca neutron-shell gap energies are compared with shell gap energies obtained from the K and Ca mass excesses in AME'03. The $^{49, 50}$Ca mass excesses have been measured and evaluated previously with good accuracies \cite{AME03_1, AME03_2}, and as a consequence, our Ca measurements do not lead to a significant improvement of the Ca $S_{2n}$ and $\Delta_{n}$ values with respect to those calculated from the mass excesses in AME'03. However, the K $S_{2n}$ and $\Delta_{n}$ values calculated with our measured mass excesses are significantly different. Owing to our $^{49}$K measurement, the two-neutron separation energy at $N=30$ is lower by about 700 keV in agreement with the falling trend observed in Ca, Sc, Ti, and V and alters the neutron-shell gap energy curve by increasing the energy at the $N=28$ shell closure by approximately 1 MeV. This observed reinforcement of the $N=28$ neutron-shell gap energy for $Z=19$ is also seen in Fig. \ref{fig:K_SGvsZ}, where the evolution of the $N=28$ shell gap energy is plotted as a function of $Z$. Combined with Jurado \textit{et al.}'s recent results in Cl \cite{jura07}, our new value for the $N=28$, $Z=19$ shell gap energy seems to indicate that the region of low atomic number when moving away from the doubly magic $^{48}$Ca, is much more stable than previously thought from the previous experimental results in K.

Fig. \ref{fig:K_SGvsZ} also includes neutron-shell gap energies predicted by various mass models, namely, the Duflo-Zuker'95 \cite{DUZU95}, Finite-Range Drop Model (FRDM'95) \cite{FRDM95}, and the  state-of-the-art microscropic Hartree-Fock-Bogoliubov \cite{HFB14, HFB17} mass models. While many more models exist, this selection is representive, as explained in detail in Ref. \cite{lunn03}. Since the comparisons in that work (performed with HFB-1 and HFB-2), considerable progress has been achieved in the predictive power of HFB models with HFB-17 reaching a root-mean-square error on known masses to better than 0.6 MeV. In Fig. \ref{fig:K_SGvsZ}, the most salient feature of all models is the $N=Z$ case of $^{56}$Ni \cite{ni56}. This is not very surprising since these are special cases (exhibiting the so-called Wigner energy (see Ref. \cite{lunn03}) for which all models have a specific, phenomenological term that is adjusted with the help of well-known masses.

The FRDM masses show a remarkable absence of the extra binding of the doubly-magic $^{48}$Ca nuclide. The results also show a small staggering that is not seen experimentally. The Duflo-Zuker results, on the other hand, show a remarkable smoothness.  This behavior reflects the continuous nature of the Duflo-Zuker formula, derived from the shell model Hamiltonian. The extra binding for $^{48}$Ca is visible, although much less pronounced than experiment and the $N=28$ shell is predicted not to be quenched. This is in marked contrast to the HFB masses that indicate a rather intense quenching. The HFB masses also correctly model the enhanced binding of the doubly-magic $^{48}$Ca.

The stronger $N=28$ shell strength for $Z=19$ is intriguing. An added binding is evident for the doubly-magic case of $^{48}$Ca but this extra stability is still largely present one proton below, as the new results attest. One question arises: Will the $N=28$ shell closure prove harder to quench farther from stability? The models studied in this paper give different predictions. If $N=28$ does quench, our results indicate that it will happen much more abruptly than previously thought. Mass measurements of lighter $N=28,30$ elements take on an obvious importance and the results of Jurado \textit{et al.} \cite{jura07} show that such measurements must be of adequate precision in order to draw any conclusions. This emphasises the importance of Penning-trap mass spectrometry.

\section{Conclusion}
Direct mass measurements of neutron-rich K and Ca isotopes have been carried out with the TITAN Penning-trap online system at TRIUMF-ISAC. The achieved precision is a factor 10-100 better than the latest Atomic Mass Evaluation (AME'03). Although, our $^{49, 50}$Ca results confirm previous measurements, we find, however, deviations of 7$\sigma$ and 10$\sigma$ for $^{48,49}$K, respectively. Our $^{49}$K mass excess lowers the two-neutron separation energy at $N=30$ by approximately 700 keV, when compared with the separation energy calculated from the mass excesses published in AME'03. Hence, the resulting neutron-shell gap energy at $N$=28 is larger than the shell gap energy calculated from the mass-excess values in AME'03 by approximately 1 MeV. Further mass measurements in neutron-rich K, Ca, and Sc are planned to gain an insight into a predicted shell closure around $N=34$ \cite{honm04}.

\begin{acknowledgments}
We show appreciation to G. Audi for comments and suggestions. This work was supported by the Natural Sciences and Engineering Research Council of Canada (NSERC) and the National Research Council of Canada (NRC). S.E. acknowledges support from the Vanier CGS program, T.B. the support from the Evangelisches Studienwerk e.V. Villigst, and V. V. S. the support from the Studienstiftung des deutschen Volkes.
\end{acknowledgments}

\end{document}